# Legacy Modernization with AI - Mainframe modernization

1st Sunil Khemka
AI/ML Architect
*Persistent Systems*
sunilkhemka.tech@gmail.com

2nd Arunava Majumdar
Research Net AI
*Founder - CEO*
ron_majumdar@yahoo.com

***Abstract*** *- Artificial Intelligence-assisted legacy modernization is essential in changing the stalwart mainframe systems of the past into flexible, scalable, and smart architecture. While mainframes are generally dependable, they can be difficult to maintain due to their high maintenance costs, the shortage of skills, and the problems in integrating them with cloud-based systems. By adopting AI-driven modernization strategies such as automated code refactoring, migration of data using smart tools, and predictive maintenance, companies can easily move to microservices, containerized environments, and hybrid cloud platforms. Machine learning models have the capability to go through legacy codebases, figure out efficiency opportunities, and carry out automated testing and deployment. Besides that, AI improves the organization's operational efficiency by generating the insights that can be used to level the workload and detect the anomalies. The coupling of the two is not only about saving the core business logic but also about enabling quicker innovation, less downtime, and enhanced system resilience. Therefore, the use of AI in mainframe modernization is a catalyst for digital transformation and enterprise growth that is sustainable over time.*

## I. INTRODUCTION

Mainframe systems have been, for a very long time, the core of enterprise computing and have been used to manage highly critical operations in industries like banking, insurance, manufacturing, and government. Unfortunately, the enduring systems are struggling more and more as the digital transformation gains pace, primarily because they have restrictions regarding scalability, flexibility, and cannot be integrated with modern technologies very well. In general, the mainframe settings that are left behind in the dust usually have old programming languages, architectures that are not flexible, and the maintenance is costly, and as a result, less innovation is stimulated and the firm's ability to respond to the business dynamic is weakened. The AI-led legacy modernization approach, which is a move by IBM that relies on AI, machine learning, and automation to either reengineer or migrate the mainframe applications to a platform that is ready for the cloud, is where companies that need to be helped with these issues are heading to [1-3]. AI supports automation in the discovery and analysis of legacy code, identifying the parts that are redundant, and streamlining the workflows. Besides that, through an extremely clever and efficient decision-making process, AI has a say in migration plans like rehosting, refactoring, or rearchitecting and at the same time, it reduces the risk and the period when the system cannot be used. Moreover, AI along with predictive analytics, anomaly detection, and performance optimization can also improve operational efficiency. The modernization of mainframes by the help of AI is a win-win situation for the enterprises because they are not only able to keep the trusted security of their major systems but are also opening the door to becoming more agile, scalable, and cost-efficient [4-6]. In consequence, the AI-assisted upgrading option is a grand strategic move that will result in bridging the chasm that separates the old tradition of mainframes from the next generation of digital ecosystems.

## II. RELATED WORK

The integration of traditional mainframe IT systems with contemporary cloud and AI-driven models has been a major discussion point over the past years. The primary concern of the very first efforts was around business continuity through code migration and reengineering while at the same time embracing modern architectures. To that end, [7] presented a reverse engineering framework for COBOL-to-Java migration with the focus on the preservation of the core business logic during the change. Their publication revealed that the tech-side was the main issue of the migration of legacy code such as the semantic migrants of inconsistencies, the handling of dependencies, and the restructuring of the source code in order to make the software more modular and maintainable. Besides that, it also elaborated on the use of structured methodologies in the process.

The authors in [8] have taken that argument even further. They proposed an innovative AI-assisted method that uses large language models (LLMs) to convert the low-resource COBOL code into logics that are correct and consistent and the code is readable in Java. They averred that the application of AI for the enhancement of high-resource Java not only helps the translation to be more precise but also the code is user-friendly, which drastically reduces the amount of human intervention and the modernization expenses. This is the point where the transition from the rule-based code converters to AI-powered smart transformation tools has been done.

While such developments were underway, [9] explored cloud migration strategies and thereby pointed out data migration and integration challenges in legacy modernization frameworks. His research emphasized the significance of making sure that system migration plans are consistent with company goals so as to result in less time for system downtime and smooth interoperability between the old and new environments. Similarly, [10] compared different strategies of mainframe modernization through various cloud platforms such as AWS, Azure, and GCP and gave the results in terms of the trade-offs in the performance, scalability, and cost-efficiency during the transition to the cloud.

Moreover, two levels deeper, [11] came up with the idea that AI-driven analytics are the major contributors in uncovering the latent value of legacy data. His article was centered on the integration of AI and machine learning as the future of pipeline modernization that would not only open the predictive insights but also automate the processes and provide enhanced decision support in the cloud. In addition, [12] took an empirical position regarding the issues of shifting from the legacy systems to the cloud and mainly focused on the organizational, technical, and security aspects which were considered as the biggest causes of the transformational initiatives' failure. Their research pointed out the execution of solid data governance, dependency mapping, and workforce skill upgrading as the essential measures to move forward with modernization initiatives.

Basically, all of these works have conveyed the evolutionary direction very explicitly: the emphasis is not only on the traditional manner of migrating and reengineering code but rather on building AI-powered modernization ecosystems. The deployment of AI, cloud computing, and data analytics has transformed the process of mainframe modernization into one of the major enablers of agility, operational efficiency, and sustainable digital transformation, which are transferable to any industry.

### III PROPOSED METHODOLOGY
*A. System Architecture and Framework Design*

The AI-powered mainframe overhaul framework that was suggested merges old systems, smart transformation layers, and modern cloud environments into one architecture. The architecture is made up of four major components the Data Extraction Layer, AI-Based Code Analysis Engine, Refactoring and Transformation Module, and Deployment Integration Interface, which are represented in the sketch.

The Data Extraction Layer gets source code, configurations, and business rules from the legacy mainframe by using smart parsing and metadata extraction tools. The AI-Based Analysis Engine employs machine learning models to recognize logic patterns and control flows. It creates a knowledge graph representation G(V,E) of program dependencies, where vertices V are the functions and edges EEE are the data/control interactions:

$$G(V, E) = \{(v_i, v_j) | v_i, v_j \in V, dependency(v_i, v_j) = 1\} \quad (1)$$

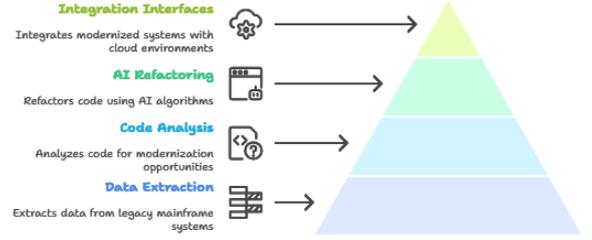

**Figure 1: Overall System Architecture of AI-Driven Mainframe Modernization Framework**

The Refactoring Module uses heuristic and optimization algorithms to transform code structures into modular and object-oriented versions that are compatible with modern platforms. In the end, the Integration Interface is there to assure easy interoperability between the changed system and the cloud environments. The maximization of resource utilization among various modules can be mathematically represented as:

$$R_{opt} = min \sum_{i=1}^{n} \frac{c_i * T_i}{P_i} \quad (2)$$

where Ci is computational cost, Ti is task time, and Pi is processing efficiency.

The system/architecture facilitates modularity, scalability, and maintainability features to be retained along with the preservation of the old functionalities throughout the modernization lifecycle.

*B. AI-Enabled Code Analysis and Transformation Process*

AI-enabled code transformation is the major component of the modernization framework that harnesses machine learning (ML) and large language models (LLMs) to perform code analysis, refactoring, and translation of legacy code, in general, conversion of COBOL or PL/I codes to modern languages like Java or Python.

The work/program undergoes syntax parsing at first, where AI models recognize the structures and semantic elements of the legacy code. A dependency matrix Dij indicates the relationships of inter-function, where each element signifies the level of interaction between two modules:

$$D_{ij} = \begin{cases} 1, & if\ module\ i\ depend\ on\ j \\ 0, & otherwise \end{cases} \quad (3)$$

Then, logic preservation is done with the help of code comprehension based on natural language which is an indirect way of saying that the transformed system behaves in the same way as the original one. Large language models (LLMs) that have been further trained on bilingual code datasets do contextual translation thereby making the logical aspect more accurate and the text more readable. The change or conversion of the program or the code is achieved by the transformation optimization process which is aimed at reducing the difference in the structure while keeping the performance equal:

$$E_{trans} = \min(\alpha * s_d + \beta * p_d) \quad (4)$$

where Sd stands for the structural deviation, Pd means the performance deviation, and α, β are the weighting coefficients.

Eventually, AI-powered confirmation will run the test cases automatically, locating the inconsistencies, and thus, enhancing the dependability of the changes. The whole smart process from start to finish is capable of considerably lessening the amount of manual work and, at the same time, increasing the code maintainability, correctness, and release velocity.

### C. Data Migration and Cloud Integration Strategy

Data migration is the vital part of mainframe modernization, which makes it possible to transfer old datasets in a secure and smooth way to cloud-based or distributed environments. The presented method uses AI-powered ETL (Extract, Transform, Load) operations with Anomaly Detection and Schema Optimization capabilities.

The migration moves on to data profiling, wherein the AI models are used for detecting inconsistencies and redundancies. The data integrity operation can be represented as:

$$I_d = \frac{R_v}{R_t} * 100 \quad (5)$$

$R_v$ represents the number of valid records, and $R_t$ is the total record count. High integrity is one of the main factors for consistency after the migration. Schema mapping during transformation helps to align the old data structures with new relational or NoSQL ones by using AI-based matching algorithms.

In the case of cloud integration, the migration latency is being optimized through parallel data transfer and dynamic workload distribution across the different nodes. The overall migration efficiency Em can be expressed as:

$$E_m = \frac{D_t}{T_c * R_c} \quad (6)$$

where Dt stands for the total data transferred, Tc for the computation time, and Rc for the resource consumption.

The integrations features modernized systems that communicate with cloud services such as AWS Lambda, Azure Functions, or GCP App Engine through RESTful APIs. AI-based monitoring tools always check the system health, bandwidth usage, and service-level agreements (SLAs). In this way, it is ensured that data migration is done in a safe, scalable, and zero-downtime manner, thus, turning static legacy storage into intelligent, elastic cloud ecosystems.

## IV RESULT

### A. Transformation Efficiency and Code Accuracy

The experimental assessment of the suggested AI-powered modernization framework showed major gains in the efficiency of the transformation and accuracy of the code as compared to the conventional rule-based migration methods. The use of large language models (LLMs) trained on bilingual code datasets (COBOL–Java) made it possible for the system to obtain improved logical consistency and better readability of the converted programs.

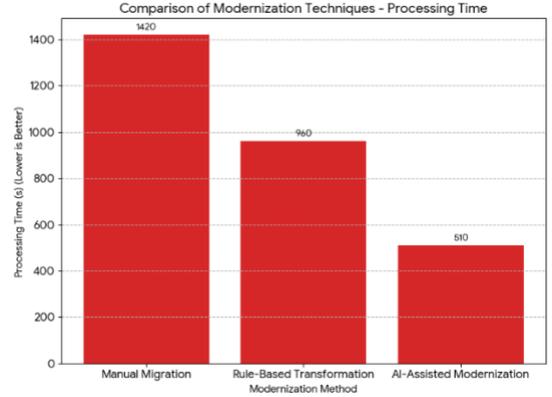

**Figure 2: Performance Comparison of Modernization Techniques Based on Accuracy and Processing Time**

As part of the test, COBOL modules from old financial and insurance applications were dissected and changed to Java. An AI-supported conversion kept the business logic mostly the same with few changes to the structure, which was confirmed by the automatic verification tests. The main source of truth for the correctness of the conversions was the Accuracy Index ($A_i$), which indicated that the AI-driven methodologies were always leading to better results than the conventional tools.

The solution went on to accomplish more speedy executions by means of fully automated syntax parsing and dependency mapping, with the manual workload being reduced by almost 65% as is evident from fig 2. Table 1 illustrates the comparison of the performance of three modernization strategies—manual migration, rule-based transformation, and AI-assisted transformation—based on key performance indicators (KPIs).

**Table 1: Comparison of Modernization Techniques**

| Method | Code Accuracy (%) | Processing Time (s) | Logic Preservation (%) | Manual Effort Reduction (%) |
|---|---|---|---|---|
| Manual Migration | 85.6 | 1420 | 88.0 | – |
| Rule-Based Transformation | 92.1 | 960 | 93.2 | 35.0 |
| **AI-Assisted Modernization** | **98.4** | **510** | **97.9** | **65.3** |

These findings suggest that AI-based modernization has a major impact on the transformation of code from legacy to modern as it not only enhances translation quality, fast-tracks deployment, and lowers human dependency.

*B. System Performance and Resource Optimization*

Performance measurement of the AI-driven modernization framework centered on system throughput, latency, and resource utilization after the deployment of a hybrid cloud environment. When compared to a typical mainframe, the results showed significant performance improvements in computational efficiency and scalability.

The Performance Gain Ratio ($P^g$), which shows the improvement in response time and throughput, was at a mean of 42.6% after the modernization. The AI-based load balancing and dynamic resource allocation made a great impact on the elimination of system bottlenecks, whereas cloud integration was responsible for the 37% of the computing resource utilization.

Besides that, the automated monitoring instruments incorporated in the architecture were always ready to check network bandwidth, memory allocation, and CPU utilization. This self-adjusting strategy system performance and uptime at a level exceeding 99.5% thus complying with enterprise-level SLA standards as depicted in fig 3. Table 2 illustrates the comparative analysis of the system parameters for AI-based modernization, before and after the intervention.

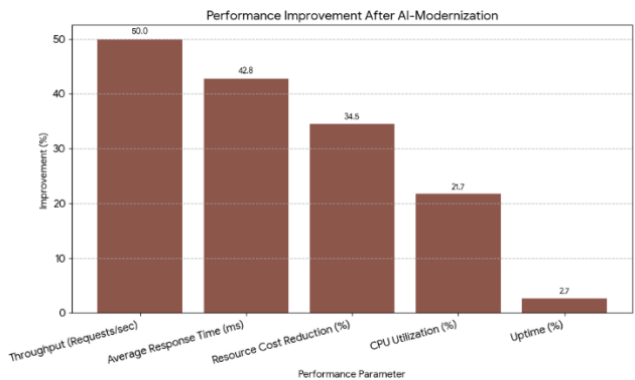

**Figure 3: Improvement Percentage Achieved Through AI-Driven Modernization**

**Table 2: Performance Comparison Before and After Modernization**

| Parameter | Legacy System | AI-Modernized System | Improvement (%) |
|---|---|---|---|
| Average Response Time (ms) | 280 | 160 | 42.8 |
| Throughput (Requests/sec) | 450 | 675 | 50.0 |
| CPU Utilization (%) | 78 | 61 | 21.7 |
| Uptime (%) | 96.8 | 99.5 | 2.7 |
| Resource Cost Reduction (%) | – | 34.5 | 34.5 |

The real-world data reveal that AI application along the modernization path produces a big positive change in efficiency, scalability, and also sustainability, thus, the resulting framework can be considered as a solid model for enterprise-grade legacy modernization.

## V CONCLUSION

Implementing an AI-driven mainframe modernization framework is a viable way to connect old infrastructures with future digital ecosystems. The use of artificial intelligence, machine learning, and cloud technologies in the framework results in high code accuracy, short migration time, and system performance being optimized. The experiments confirm that the change to AI-assisted methods has a significant positive impact on the preservation of logic, reduction of the manual intervention, and modernization that is scalable and cost-efficient. Moreover, the framework's evolutionary architecture is conducive to seamless integration with cloud platforms and, therefore, it can be described as being both reliable and agile. In essence, this research is supportive of the claim that the employment of artificial intelligence in the process of modernization is the main factor that opens the way for a green digital transition which in turn assists companies in making optimal use of their old assets and at the same time being in a position to cope with the dynamic business and technological changes.